\documentclass[preprint2]{aastex}

\usepackage[]{natbib}

\shorttitle{A. A. Deshpande and J. M. Rankin}
\shortauthors{Pulsar Emission Mapping}

\begin{document}

\title{ Pulsar Magnetospheric Emission Mapping:\\
        Images and Implications of Polar-Cap Weather}

\author{Avinash A. Deshpande} 
\affil{Raman Research Institute, Sadashivanagar, Bangalore 560080 INDIA}
\and
\author{Joanna M. Rankin}
\affil{Physics Department, University of Vermont, Burlington, VT 05405 USA}

% The abstract environment prints out the receipt and acceptance dates
% if they are relevant for the journal style.  For the aasms style, they
% will print out as horizontal rules for the editorial staff to type
% on, so long as the author does not include \received and \accepted
% commands.  This should not be done, since \received and \accepted dates
% are not known to the author.

\begin{abstract}
The beautiful sequences of ``drifting'' subpulses 
observed in some radio pulsars have been regarded 
as among the most salient and potentially instructive 
characteristics of their emission, not least because 
they have appeared to represent a system of subbeams  
in motion within the emission zone of 
the star.  Numerous studies of these ``drift'' 
sequences have been published, and a model of their 
generation and motion articulated long ago by Ruderman 
\& Sutherland (1975); but efforts thus far have
failed to establish an illuminating connection 
between the drift phemomenon and the actual sites
of radio emission.  Through a detailed analysis of a
nearly coherent sequence of ``drifting''  pulses from 
pulsar B0943+10, we have in fact identified a 
system of subbeams circulating around the magnetic 
axis of the star.  A mapping  technique, involving 
a ``cartographic'' transform and its inverse, 
permits us to study the character of the polar-cap 
emission ``map''  and then to confirm that it, in 
turn, represents the observed pulse sequence.  On 
this basis, we have been able to trace the physical 
origin of the ``drifting-subpulse''  emission to a 
stably rotating and remarkably organized 
configuration of emission columns, in turn traceable 
possibly to the magnetic polar-cap ``gap'' region 
envisioned by some theories.  
\end{abstract}

% The different journals have different requirements for keywords.  The
% keywords.apj file, found on aas.org in the pubs/aastex-misc directory, 
% contains a list of keywords used with the ApJ and Letters.  These are 
% usually assigned by the editor, but authors may include them in their 
% manuscripts if they wish. 

\keywords{MHD --- plasmas --- pulsars: general, individual (B0943+10) --- radiation mechanism: nonthermal }

% That's it for the front matter.  On to the main body of the paper.
% We'll only put in tutorial remarks at the beginning of each section
% so you can see entire sections together.

% In the first two sections, you should notice the use of the LaTeX \cite
% command to identify citations.  The citations are tied to the
% reference list via symbolic KEYs.  We have chosen the first three
% characters of the first author's name plus the last two numeral of the
% year of publication.  The corresponding reference has a \bibitem
% command in the reference list below.
%
% Please see the AASTeX manual for a more complete discussion on how to make
% \cite-\bibitem work for you.   

\section*{}

Since their discovery by \citet{dcj68}, ``drifting'' subpulses 
have fascinated both observers and theorists and have often 
been regarded as one of the ``keys'' to understanding  
pulsar radio emission.  Without any adequate model, early 
studies of prominent ``drifters'' ({\it e.g.}, \citet{tah71}; 
\citet{bac73}) focussed on delineating the drift phenomenon.  
Then, Ruderman \& Sutherland's (1975) model not only 
articulated a qualitative picture of ``drifting'' subpulses 
as a system of ``carousel beams'' circulating around the 
magnetic axis within the (magnetic) polar emission region, 
but they also attributed the rotation physically to 
{\bf{E}}$\times${\bf{B}} drift and estimated how long 
these circulation times would be.  

While fascinating and theoretically compelling, 
the phenomenon has proved difficult to interpret physically.  
No such sequence has appeared precise enough to determine 
the structure and dynamics of the beams, and thus to see if 
it follows any pattern which might or might not confirm 
the expectations of the aformentioned model.  
Any sightline to a star can cross the periphery of its 
polar emission cone at only one or two points, and it 
was difficult to fix accurately just where these points 
fell in relation to the rotational and magnetic axes.  
Nor was it clear to what extent the ``drift'' rates 
were aliased by the rotation frequency of the star.  
Within the Ruderman \& Sutherland model, the total number 
of subbeams must be ascertained, or no calculation of 
the circulation time is possible---and thus no quantitative 
connection can be made with theories of the polar-cap 
``gap'' region.   It is a telling statement on this 
unfortunate state of affairs that the most extensive 
studies of a pulsar with a highly regular ``drifting''-subpulse 
pattern, B0809+74, were published 25-30 years ago! [\citet{tay71}; 
\citet{man75}].

Pulsar B0943+10 is another pulsar in the same class and is remarkable 
in other ways as well.  One of a handful of pulsars discovered below 
300 MHz \citep{vit69}, it weakens markedly at higher frequencies and 
has never been detected above 600 MHz.  Its profile evolution (and 
``drifting'' subpulses) identify it as a member of the conal single 
($S_d$) class \citep{jr93b}, and we now know that its steep
spectrum is due to a sightline traverse so peripheral that it ultimately
misses the ever narrower high frequency cone.  The pulsar's long sequences 
of ``drifting''  subpulses have prompted several different studies 
[\citet{tah71,bac75,sao75}]---all showing strong fluctuations at a 
frequency of 0.46 cycles per rotation period (hereafter, c/$P_1$) and a 
weaker feature near 0.07 c/$P_1$.  Two profile ``modes'' also have been 
identified, a ``B'' (for ``bright'') mode in which subpulse ``drift'' 
is very prominent, and a ``Q''  (for ``quiescent'')  mode wherein the 
subpulses are weaker and disorganized [\citet{sul84,sul98}].  
For all these reasons the pulsar's subpulse sequences are difficult 
to observe, and our study is based on 430-MHz observations benefiting 
by the high sensitivity of the large Arecibo telescope in Puerto Rico.

\begin{figure}
\figurenum{1}
\epsscale{1.}
\plotone{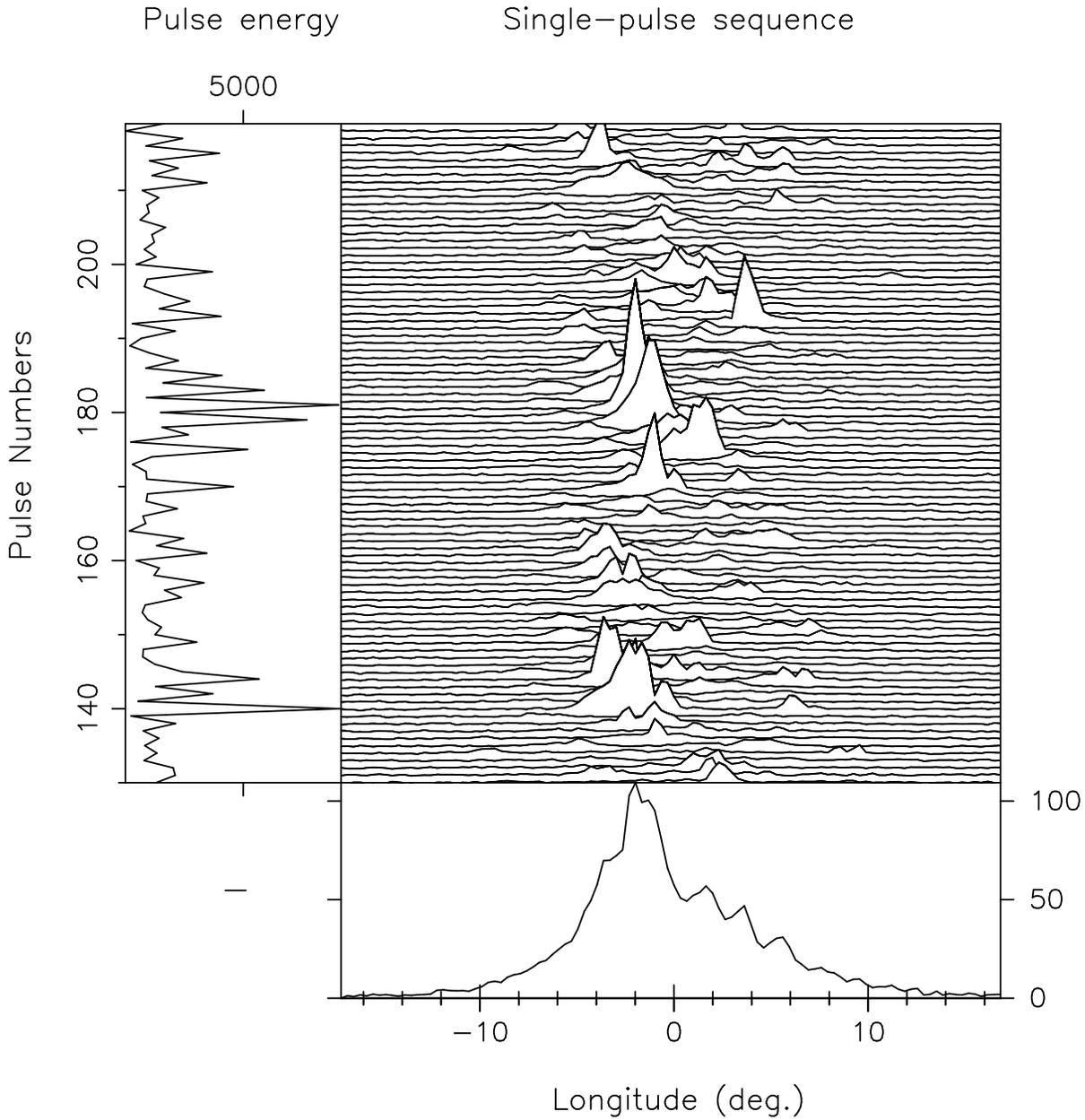}
\caption{A series of individual pulses from pulsar B0943+10 (centre
panel) along with their average (bottom panel) and energy (left
panel) as a function of pulse number and longitude (360$^o$
longitude correspond to one stellar rotation).  Note the
``drifting'' subpulses, the alternate-pulse modulation, and
the ``single'' average profile.\label{fig:fig1}}
\end{figure}

\begin{figure}
\figurenum{2}
\epsscale{1.}
\plotone{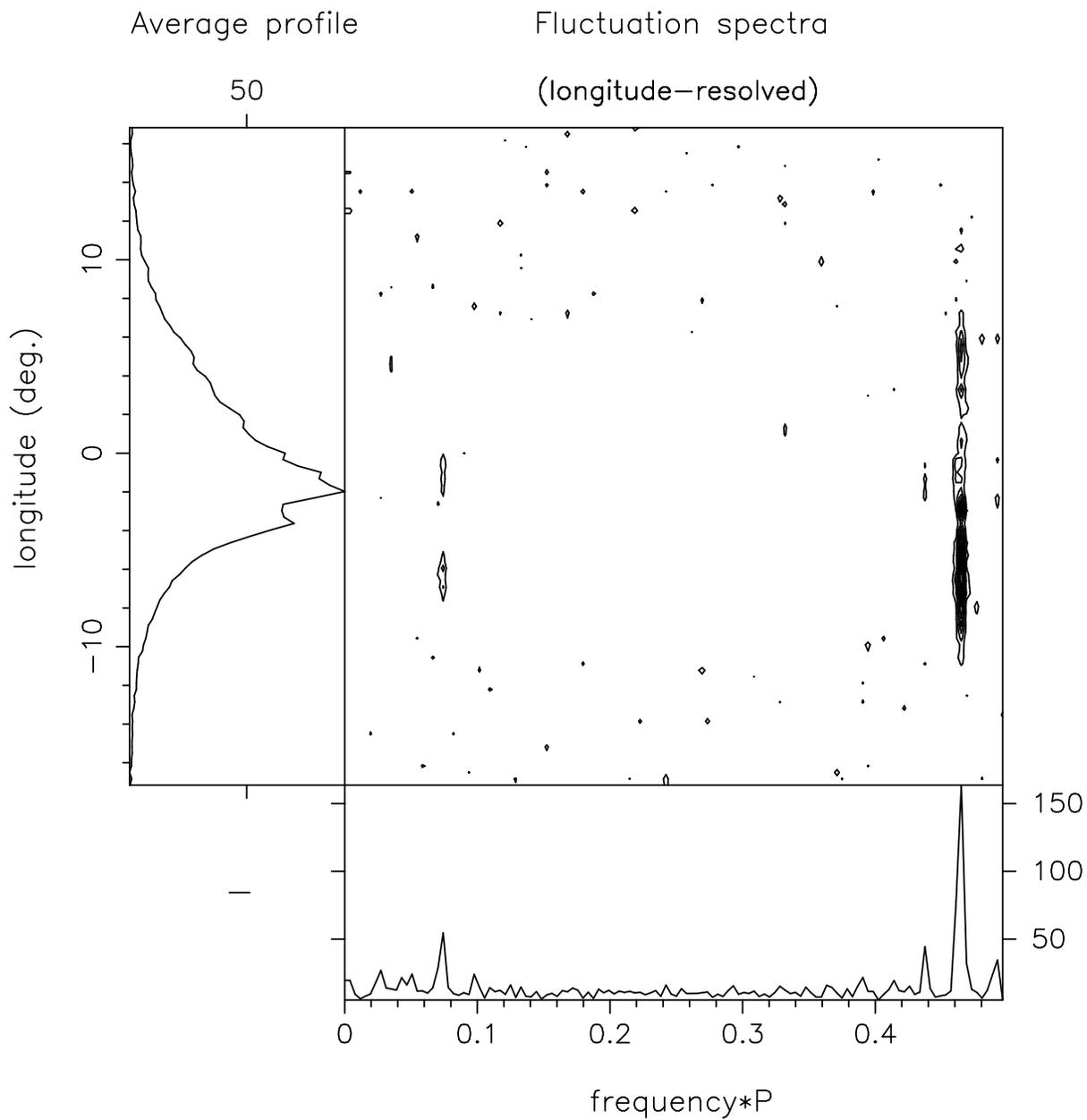}
\caption{Fluctuation-spectral power (centre panel)
as a function of longitude and frequency as well as the integral spectrum
(bottom panel).  Note the primary and secondary features at
about 0.46 and 0.07 cycles/period as well as the symmetrical ``sidebands''
around the former.\label{fig:fig2}}
\end{figure}

\begin{figure}
\figurenum{3}
\epsscale{1.}
\plotone{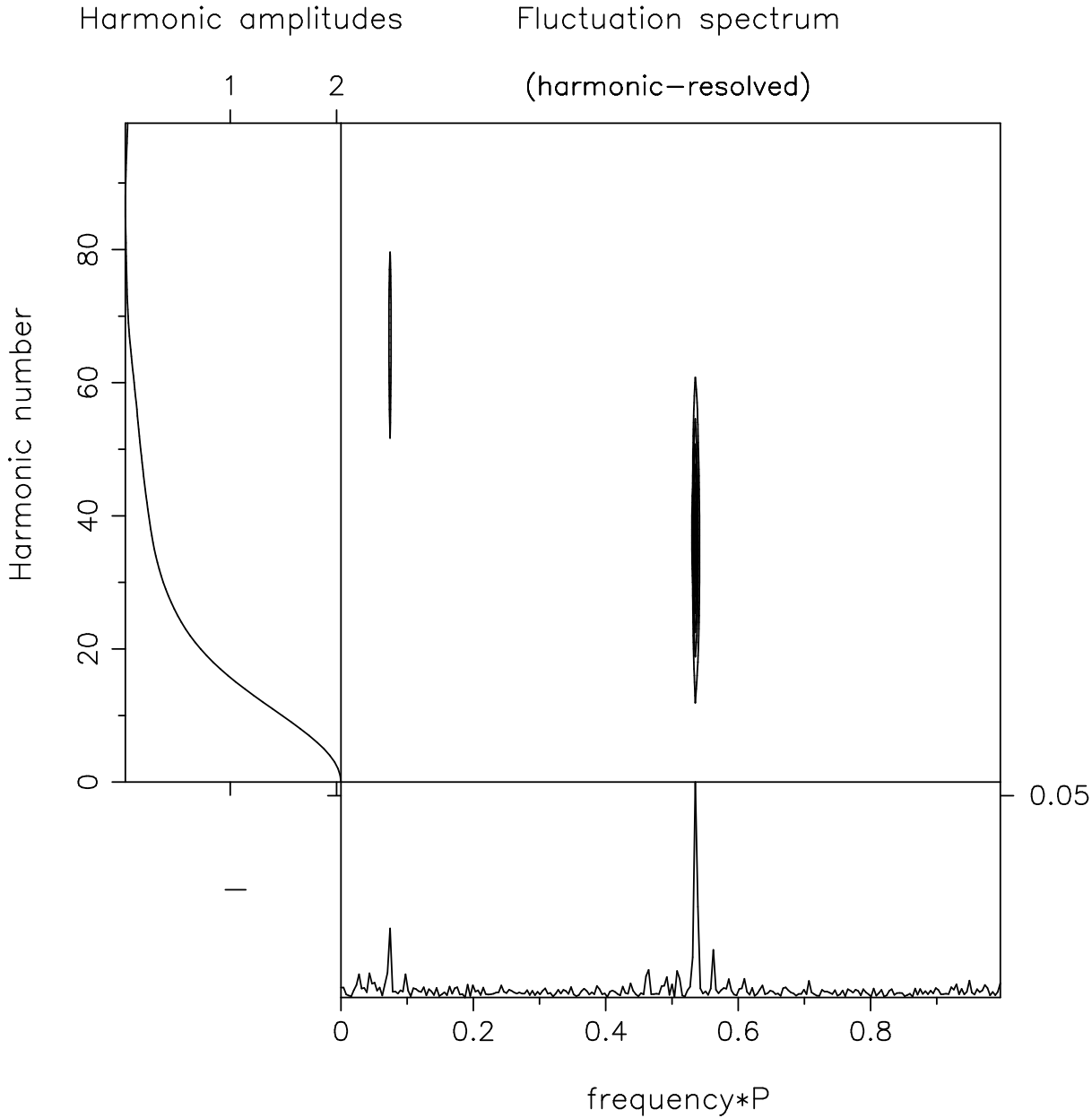}
\caption{Fluctuation spectrum of a continuously-sampled time sequence
reconstructed from the (gated) single-pulse sequence used in Fig. 2.
The spectrum is presented in a matrix form with
each row showing a section of the spectrum between $n/P_1$ to
$(n+1)/P_1$ for harmonics $n=0,1,2,...$ of the rotation frequency $1/P_1$.
The harmonics at $n/P_1$ are shown separately (left panel) and
are essentially the Fourier components of the average profile.
All other frequency components (up to $100/P_1$) are
given as a contour plot (centre panel), and the column sum of these
components, collapsed into a $1/P_1$ interval (bottom panel),
can be compared directly with Fig. 2.
Note that the principal feature now falls at an
(unaliased) frequency of about 0.535 c/$P_1$.  Note also that whereas the
Fourier amplitudes of the principal feature peak at about harmonic (row)
number 35 (corresponding to a $P_2$ value of some $10.5^o$), those
of the secondary feature peak around number 70---thus
demonstrating their harmonicity.\label{fig:fig3}}
\end{figure}

A short sequence of ``B''-mode pulses is given in Figure~\ref{fig:fig1},
together with their average profile, and the 256-pulse fluctuation 
spectra in Figure~\ref{fig:fig2} exhibits both the 0.46 and 0.07 c/$P_1$ 
features very clearly.  The former is hardly resolved, whereas the 
latter appears to be just so.  A harmonic connection between these 
two features has been debated [\citet{tah71,bac75,sao75}],
but never clearly determined because of the possible ambiguities 
involving aliases in an ordinary ``longitude-resolved''  
fluctuation spectrum, such as the one given in the figure.  We have 
resolved the matter by exploiting the fact that the associated 
modulation is {\it continuously} sampled within the finite duration of 
each pulse. Therefore, we first Fourier transformed the entire 
pulse sequence, and then, from this unfolded spectrum, established 
that the sets of sidebands associated with the features were harmonic.
This harmonic-resolved spectrum is shown in Figure~\ref{fig:fig3}, and 
will be discussed more fully in a subsequent paper \citep{dar99}.  
It can now be confidently asserted that the true 
{\it phase}-modulation frequency is 0.5352$\pm$0.0006 c/$P_1$ (though 
seen at 0.46 c/$P_1$ as a first-order alias) and that the secondary 
feature at 0.0710$\pm$0.0006 c/$P_1$ (a second-order alias of 1.07 c/$P_1$) 
represents its second harmonic. These frequencies vary slightly 
with time and are thus slightly different in our three observations, 
but always fall close to the 1992 October values given above. 

With the aliasing resolved, it is also clear that the time interval between 
``drift'' bands---usually denoted by $P_3$---is just 1/(0.5352 c/$P_1$) or some
1.87 $P_1$/cycle. Thus, just less than two rotational periods are required 
for subpulse emission to reappear at the same longitude, both giving the 
sequence its strong odd-even modulation and establishing that the actual, 
physical subpulse motion is from trailing to leading (or right to left) 
in the diagram---that is, {\it negative}, or in the direction of 
decreasing rotational longitude.  

\begin{figure}
\figurenum{4}
\epsscale{1.}
\plotone{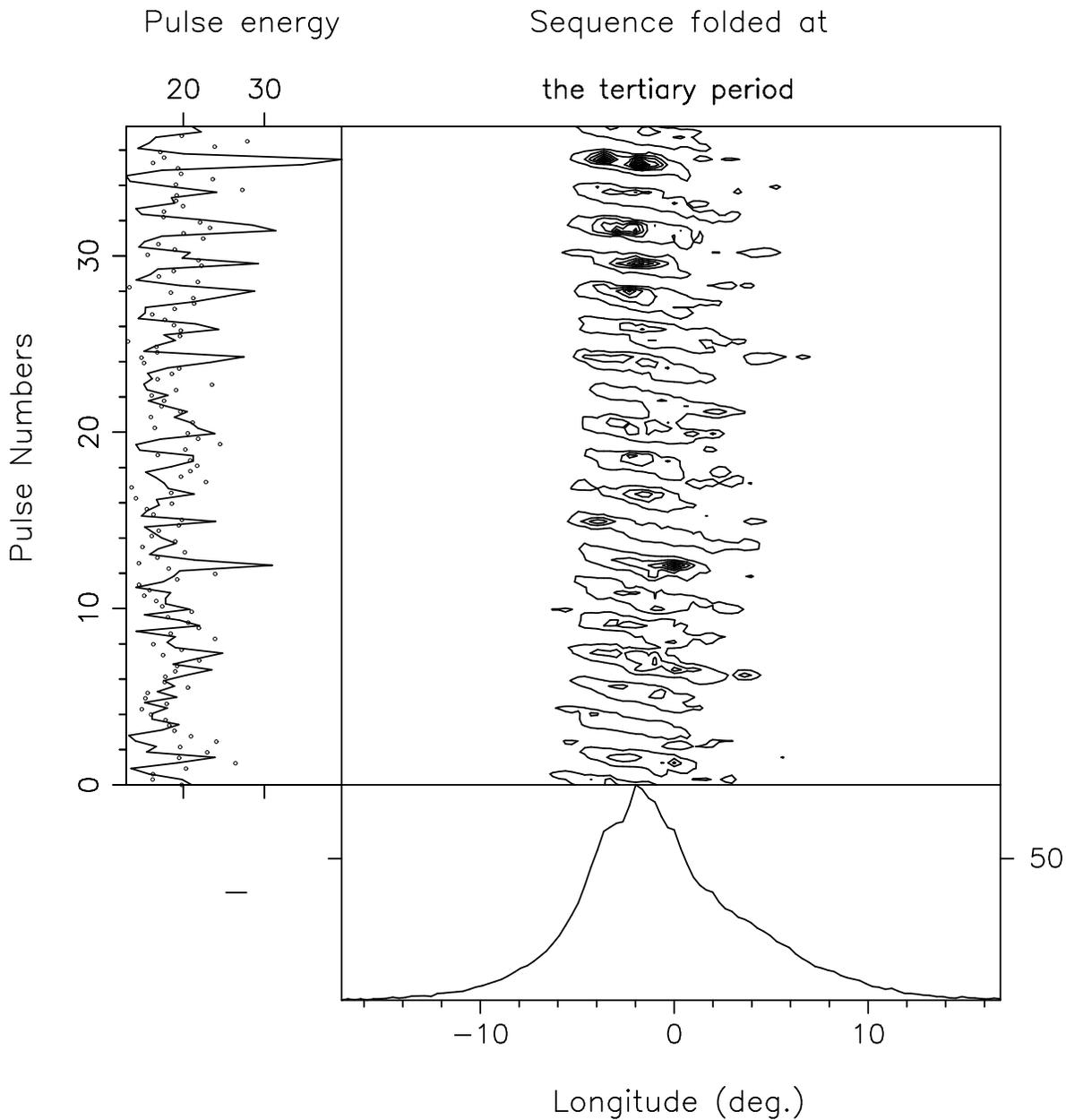}
\caption{Pulse sequence folded at the full 37.35-period modulation cycle
(centre panel).  Twenty emission centres can be readily identified,
and their relative intensity is shown as a function of longitude
(bottom panel) and modulation phase in units of pulse number (left
panel).  For comparison, the dots in the latter show the result
using a tertiary period that is 2\% different than the correct one.
\label{fig:fig4}}
\end{figure}

The fluctuation spectra of Figs.~\ref{fig:fig2} \& \ref{fig:fig3} also show a pair of 
symmetrical sidebands associated with the primary feature.
These sidebands fall 0.027 c/$P_1$ higher and lower than the principal 
fluctuation and represent an {\it amplitude} modulation {\it on}
the phase modulation.  This circumstance, viewed
along with the remarkable stability of the phase fluctuation, 
demands that these features also have an harmonic relationship 
with the fundamental fluctuation frequency, and so we are not 
surprised to find that (0.535 c/$P_1$)/(0.027 c/$P_1$) = 20.01$\pm$0.08, 
which is an integer, 20, to well within the errors.  The entire 
modulation cycle is then just 20 times the 1.87-period one, or 
some 37.35$\pm$0.02 periods, and this circumstance is beautifully 
confirmed in Figure~\ref{fig:fig4}, where the full sequence has been folded at 
this period. 

We have also studied the geometry and polarisation of the pulsar 
in unprecedented detail.  All the available mean profiles (seven 
frequencies between 25 and 430 MHz) were used to model the star's
conal beam geometry---thus fixing the angles that the magnetic
axis and sightline make with the rotation axis, respectively.  
Four different configurations were explored, corresponding to both 
inner and outer cones [\citet{jr93a,jr93b,mit99}] and to poleward 
and equatorward sightline traverses, respectively.  All show that at 
430 MHz the sightline makes an exceedingly tangential traverse through the 
emission cone---at close approach still some 102\% of the half-power 
radius of the cone.  Only a poleward traverse could be reconciled with 
both the observed polarisation-angle sweep rate of some -3$^o/^o$ 
\citep{sul98} and the requisite magnetic azimuth interval ($18^o$) 
between adjacent subpulses,\footnote{The inner-cone solution 
corresponds to $\alpha$ and $\beta$ some $11.6^o$ and $-4.3^o$,  
respectively, and all of our calculations are based on these values. 
Other solutions such that $\sin\alpha/\sin\beta \approx -3^o/^o$ 
only change the overall scale of the map.} which then also 
fixes the actual, physical sense of rotation of the ``nearer'' 
(rotation) axis as clockwise.  We calculate the magnetic azimuth 
interval between adjacent subpulses on the basis of a) their 
longitude separation and b) the polarisation-angle rotation between 
them \citep{rad69}, obtaining values of 18$^o$ independent of (inner 
or outer) conal radius.  Clearly, entities at 18$^o$ intervals are 
20 in number, implying a vigesimal beam system circulating in 
magnetic azimuth.

Pulsar researchers have long tended to view ``drifting''  subpulses 
as a pattern of ``carousel'' beams [\citet{ras75,oas76}], rotating 
around the magnetic axis, but so far it has not been possible to 
verify this picture either by tracing a given beam through its circuit 
or by establishing the total number of beams.  We see that our 
observations and analysis of pulsar 0943+10 now give clear 
signatures of 20 beams, rotating counterclock-wise around its 
``emitting'' magnetic axis in a total time $\hat{P_3}$ of some 37  
periods or 41 seconds.  Furthermore, we understand the pulsar's emission
geometry sufficiently well that we can transform the received sequence
from the observer's frame into a frame which is rotating about the
magnetic axis of the pulsar.  This ``cartographic''  transform then maps
the intensity as a function of longitude (with respect to that of the 
magnetic axis) $\phi-\phi_o$  and pulse number $k-k_o$ into magnetic-polar 
colatitude $R$ and azimuth $\theta$ (rotating with period $\hat{P_3}$).  
Successive pulses then sample the intensity along successive chord-like 
traverses (corresponding to different azimuth intervals) through the 
rotating-beam pattern. 

\begin{figure}
\figurenum{5}
\epsscale{1.}
\plotone{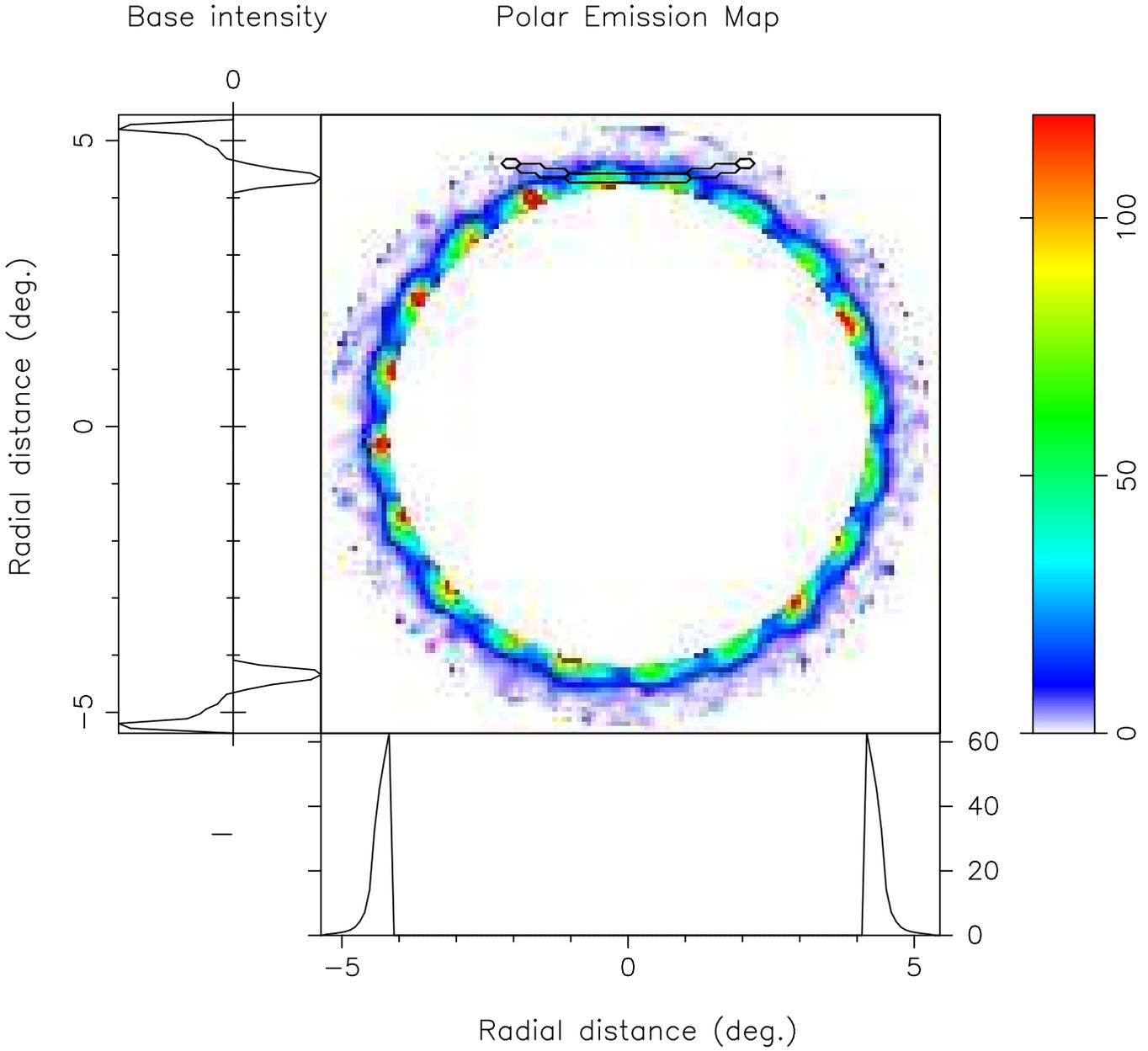}
\figcaption{Results of the ``cartographic'' transformation. The individual-pulse
sequence has been mapped onto a rotating frame centred on the
magnetic axis.  Azimuthal average and minimum (unfluctuating ``base'')
intensity as a function of magnetic colatitude are given in the bottom
and left panels, respectively.  Note the twenty emission centres and
their somewhat elongated shape. A portion of the sightline traverse
is also indicated at the top of the figure.\label{fig:fig5}}
\end{figure}

Applying this cartographic transformation to the ``B''-mode portion  
of our 1992 observations, Figure~\ref{fig:fig5} shows the average 
configuration of the 20 subbeams which produce the ``drift'' sequence.
The rotation of adjacent beams through our sightline produces the 
secondary {\it phase} modulation, and the periodic pattern of varying 
beam intensities gives it its tertiary {\it amplitude} modulation.  
We find that the subbeams vary, both in intensity and in azimuthal spacing, 
over 100-second time scales.  Closely spaced beams
tend to be weaker, and some beams are observed to bifurcate temporarily, 
but the 20-fold pattern is always maintained strongly as a stable 
configuration. The elongated shape of the beams can be understood
as a result of truncation, as most of the emission falls inside of
the 430-MHz sightline traverse.  Indeed, a similar map at a much lower 
frequency (111.5 MHz), where we expect the sightline to sample emission
interior to that one above, shows a more nearly circular subbeam shape.

How sensitive are the conclusions drawn from the cartographic transform 
to any assumptions in the analysis?  The mapping procedure is exquisitely 
sensitive to the circulation time, and only a little less so to the 
longitude of the magnetic axis, to the polarisation-angle sweep rate, 
and to the angles specifying the colatitude of the magnetic axis and 
sightline ``impact'' angle.  Overall, these parameters scale 
the map---or utterly distort it.  However, the forward transform used 
to construct the maps has a true inverse transform, and this inverse 
cartographic transform can be used to ``play back'' a map in order to produce 
an artificial pulse sequence.  Only when this artificial sequence correlates 
in full with the observed sequence do we take the map as correct. 
Indeed, we have carried out such inverse transforms iteratively in a kind 
of ``search mode'' to refine each of the requisite parameters, and a 
completely consistent picture is found in observations from
1974 \& 1992 at 430 MHz and from 1990 at 111 MHz.\footnote{Moving 
images of these sequences can be viewed at sites with 
the following URL addresses: http://www.rri.res.in/$\sim$desh and 
http://www.uvm.edu/$\sim$jmrankin}

The overall structure of 0943+10's radio beam
provides new insight into the physical nature of pulsar emission.  Its
20 subbeams rotate almost rigidly, maintaining their number and spacing 
despite perturbations tending both to bifurcate a given beam and to merge 
adjacent ones---and we find an almost identical subbeam configuration 
at the two frequencies.  These circumstances suggest that the subbeam 
emission region lies along bundles of magnetic field lines 
(or ``plasma columns'') having 
``feet'' within a certain annulus on the polar cap, wherein charges are 
accelerated at some distance above the stellar surface and radiation 
occurs at progressively higher altitudes for lower and lower 
frequencies.  We can thus peer at the activity at the ``feet''  of
the plasma columns and possibly monitor changes viewable at intervals 
of the circulation time.

The subbeam radiation we observe, estimated to be emitted at some 
100-300 km height, then reflects processes occurring at lower altitudes 
along these ultra-magnetized plasma columns---indeed, some fully down 
{\it at} the surface where the electric fields caused by the star's 
rotation appear.  The ``foot'' of each subbeam column moves around 
the magnetic axis, so we cannot appeal to any surface features or 
fixed ``hot spots'' as their cause.  The entire rotating-subbeam pattern, 
averaged over time, represents a hollow-cone beam of emission,
which is emitted along the magnetic-field direction at a particular height
for a given frequency. Signatures of such conical beams are encountered 
in most pulsars, and accurately reflect the angular dimension of the 
magnetic polar cap \citep{jr93a}.  

Nonetheless, we have little guidance about where the ``feet'' of these 
plasma columns will fall on the polar cap.  Clearly, the radius 
of the active polar cap $r_{PC}$  scales as the square-root of the 
star's radius to that of the light cylinder, 
$r_{PC} = \epsilon r_*(2\pi r_*/cP_1)^{1/2} \sim 145\epsilon P_1^{-1/2}$ 
meters (assuming $r_* \approx 10$ km), and values of $\epsilon$  near unity or 
so have been taken to describe the outside edge of the emission pattern.{\footnote 
{{\it e.g.}, $\epsilon$ is $(2/3)^{3/4}$ in Ruderman \& Sutherland (1975) and 
1.5 in \citet{jra90}, 
respectively}}  Therefore, polar-cap features corresponding to the   
subbeam columns would have separations of some 45$\epsilon$ meters and 
individual sizes of some 20$\epsilon$ m across.  These dimensions 
would increase only a little even if the ``feet'' of the plasma 
columns happen to be a few stellar radii away from the star's surface.  
Interpreted another way, our estimate of the angular width of the subbeams
(based on the magnetic colatitude spread seen in the map at 111.5 MHz)
suggests Lorentz factors $\gamma > 200$.
We note again that our extremely tangential sightline permits us to sample
only the outermost portion of each subbeam at 430 MHz, and so only this 
outer portion can be mapped down onto the polar cap.
                                 
%Because our 1992 observations were well calibrated against a standard source, 
%we can readily estimate the total flux $S$ radiated by the entire beam
%pattern as about 40 Janskys.\footnote{The average flux that this 
%represents is slightly smaller than the tabulated value (\citet{tay93}),  
%probably due to the effects of scintillation.} The total radio luminosity  
%is then $L = Sd_{kpc}^2\Delta \nu$, where $d_{kpc}$ is the distance in 
%kiloparsecs and $\Delta \nu$ the bandwidth.  If $\Delta \nu \sim$ 1 GHz, 
%$L\sim 7$$\times$$10^{30} d_{kpc}^2$ ergs/s, assuming the second polar-cap 
%region emits equally.  The spindown luminosity is estimated to be some 
%1$\times$10$^{32}$ ergs/s, making the star's radio efficiency $\ge 14 d_{kpc}^2$\%.  
%On the one hand, well more than half the total luminosity may lie 
%inside our 430-MHz sightline traverse, but then, there can be some 
%doubt that the pulsar is really as far as the 0.98 kpc currently 
%estimated (\citet{tay93}; \citet{tac93}).  Nonetheless, this is a remarkably 
%high efficiency, unless the distance estimate is in error by a 
%significant factor. 

Ruderman \& Sutherland (hereafter, R\&S), nearly 25 years ago, 
identified the ``drift''-associat-ed subbeams with electrical 
breakdowns (or ``sparks'') in the polar-cap ``gap'' region, and 
argued that their circulation was due to {\bf{E}}$\times${\bf{B}} 
drift.  Although the major standpoint of their model---that even the 
enormous electric fields that must be generated would be inadequate
to overcome the binding energy of the positive charges (like
iron ions) at the neutron-star surface---now seems untenable 
\citep{pbj85,pbj86}, closely related models are still in active 
discussion [{\it e.g.}, \citet{uam95,zaq96,zha97}].
It is noteworthy that no completely independent model of 
subpulse ``drift'' seems to exist, and our maps do suggest a 
qualitative picture something like the model R\&S envisioned.  
Any subpulse ``drift'' model needs to explain first why 
subpulse-scale structures occur and then why they circulate in 
azimuth.  According to Ferraro's Theorem \citep{fap66} the 
open-field region above the polar cap must rotate with, but 
perhaps more slowly than, the star (so that, for an inertial 
observer, they would both rotate in the {\it same} direction); thus, 
in the pulsar's frame the regions will rotate oppositely---as 
is observed for 0943+10.\footnote{That the polar-cap ``drift'' 
lags the star's rotational speed may, then, be evidence for a 
``gap'' somewhere between the emission region and the stellar 
surface, but it gives no indication about whether positive 
(negative) charges would be accelerated outward---or, in 
R\&S's terms, whether the star is a {\it pulsar} or an 
{\it antipulsar} \citep{rud76}. }   

Two quantitative features of the R\&S model are noteworthy.  First, 
their ``gap'' height is some 50 m---which is also related in the 
model to the scale over which an active "spark" or 
plasma column inhibits the formation of others---and 
the 45$\epsilon$ m we infer is fully consistent 
with this value.  Secondly, the observed circulation time 
can also be reconciled with their model. $\hat{P_3}$, at 
polar-cap radius $r_s$, for drifting at velocity 
{$\bf{E}$$\times$$\bf{B}$}/$B^2$, can be written as 
$\hat{P_3} = 2\pi r_s B/$$<$$E$$>$$ P_1$, where $B$ and 
$<$$E$$>$ are the magnetic field and the average radial 
electric field in the ``gap'' region.  $<$$E$$>$  
in turn depends on the ``gap'' potential 
drop $\Delta V$ and the ``spark'' distance from the edge of 
the cap, so that the average field is perhaps some 
$<$$E$$>$$ = \Delta V/2(r_{PC}-r_s)$ [{\it c.f.}, R\&S's eq.(30)].  
Assuming that the pulsar's magnetic field is correctly computed 
as a surface value [see \citet{jan88}'s eq.(1)], then $B$ is 
2$\times$10$^{12}$  Gauss, $P_1$ 1.1 seconds, and thus $\hat{P_3}$ 
can be as large as 37 $P_1$ only if $(r_{PC}-r_s) \sim r_s$
and $\Delta V/\epsilon^2$ is well less than 10$^{12}$ V.  In 
any case, the $<$$E$$>$ in the acceleration region is required 
to be $\sim 3$$\times$$10^9 \epsilon$ Volts/meter, for the 
circulation to be due to {\bf{E}}$\times${\bf{B}} drift.  

Our maps suggest that the angular velocity associated with the 
circulation is nearly constant across the radial extent of the
subbeams, because no azimuthal shear is observed within the 
subbeams.  The exact origin and significance of any possible 
dependences of $\hat{P_3}$ on radius and other parameters 
needs to be understood, and similar estimates of circulation 
time in other pulsars (with different rotational parameters 
and geometries) should help to define this issue.  The 
finely-tuned stability of the pattern observed in this pulsar 
has more to tell.  It is unlikely that the remarkably periodic 
and stable arrangement would be possible if the required 
spacing of subbeams were not an integral submultiple of the 
circumference of the ring on which they appear to arrange 
themselves.  Pulsar B0943+10's particular characteristics may 
therefore result from a critical combination of 
parameters---namely, its rotation period, emission geometry, 
magnetic field strength and subbeam spacing, which in turn 
could be determined by the ``gap'' height.  It should not, 
therefore, be surprising to find that this remarkable 
configuration is indeed an unusual one. 

Nevertheless, averaged over many circulation times, 0943+10's 
emission pattern conclusively demonstrates the existence of the 
hollow-conical emission beams long attributed to many pulsars  
through less direct means.  The subbeam columns must be nearly 
axially symmetric with respect to the magnetic axis, because 
any significant deviation would require a contrived situation 
to reproduce the periodicities and polarization that we observe.  
Models in which only a part of the polar cap is active [{\it 
e.g.}, \citet{aas79,aro79}] are incompatible with 
these results.  

The central conclusion to be drawn from our subbeam mapping 
of pulsar B0943+10 is that the emitting pattern---which is 
so very stable over hundreds of seconds or many circulation 
cycles---is frozen neither to the fields nor to the stellar 
surface.  We must then ask where, physically, the ``memory'' 
of the subbeam pattern is carried in this system of charges 
and particle currents despite the circulation.  The detailed 
structure that we have mapped traces the activity at the 
``feet'' of the emission columns and represents the first 
direct measurements of some of the parameters---such as the 
locations, size and movement of the underlying 
pattern---reflecting the electrodynamics not very far from 
the stellar surface.  The overall continuity of the emission, 
implicit in the stability of the pattern over many circulation 
periods, demands a remarkable ``collective steadiness'' at 
every stage of the emission process---such as, generation and
acceleration of particles and {\it coherent} amplification.

In summary, pulsar 0943+10 exhibits an exqui-sitely
stable ``drifting''-subpulse pattern in its ``B'' mode sequences with
a nearly even-odd fluctuation frequency of about 0.53 c/$P_1$.  Upon first 
establishing the aliasing-order of this feature, and then accurately 
modeling the star's emission geometry, the two ``sideband'' features were
found to result from a tertiary amplitude modulation on the secondary phase 
modulation. The remarkable precision of the modulation 
rates argues for a system of subbeams circulating around the magnetic
axis, prompting the use of a novel ``cartographic''  transform to map 
the subbeam structure responsible for the pulse-to-pulse fluctuations.  
The detailed parameters of the circulating subbeams allow quantitative
assessment of the possible interpretation in terms of the system of
rotating ``sparks'' on the polar cap as suggested in the Ruderman 
\& Sutherland model. The current analysis holds considerable promise 
for studying the emission properties in other pulsars and for
assessing physical emission theories.

\acknowledgments

We thank Vera Izvekova, Svetlana Suleymanova \& N. Rathnasree for 
help with the observations, and V. Radhakrishnan \& Rajaram 
Nityananda for their insightful, helpful comments and many useful 
suggestions for improving the manuscript.  We are also grateful 
to Malvin Ruderman, Alice Harding and Jonathan Arons for useful 
discussions.   Portions of this work were carried out under US 
National Science Foundation Grants AST 89-17722 and INT 93-21974.  
Arecibo Observatory is operated by Cornell University under 
contract to the US National Science Foundation.

Correspondence and requests for materials can be addressed to either 
of the authors (email: desh@rri.res.in and rankin@physics.uvm.edu).  

\onecolumn

%\newpage

%\newpage

%\newpage

%\newpage

\begin{thebibliography}{}
\bibitem[Arons(1979)]{aro79} Arons, J. 1979, Space Sci. Rev., 24, 437
\bibitem[Arons \& Scharlemann(1979)]{aas79} Arons, J. \& Scharlemann, E. T. 1979, 
        \apj, 231, 854
\bibitem[Backer(1973)]{bac73} Backer, D. C 1973, \apj, 182, 245
\bibitem[Backer et al.(1975)]{bac75} Backer, D. C., Rankin, J. M., \& 
        Campbell, D. B.  1975, \apj, 197, 481
\bibitem[Deshpande \& Rankin(1999)]{dar99} Deshpande, A. A. \& Rankin, J. M. 1999, 
        in preparation.  
\bibitem[Drake \& Craft(1968)]{dcj68} Drake, F. D., \& Craft, H. D. jr
        1968, Nature, 220, 231
\bibitem[Ferraro \& Plumpton(1966)]{fap66} Ferraro, V. C. A., \& Plumpton, C. 
	1966, {\it Magneto-Fluid Mechanics} (London: Oxford University Press), 23.  
\bibitem[de Jager \& Nel(1988)]{jan88} de Jager, O. C., \& Nel, H. I.
        1988, \aap, 190, 87
\bibitem[Jones(1985)]{pbj85} Jones, P. B. 1985, \prl, 55, 1338
\bibitem[Jones(1986)]{pbj86} Jones, P. B. 1986, \mnras, 218, 477
\bibitem[Manchester et al.(1975)]{man75} Manchester, R. N., Taylor, J. H., 
        \& Huguenin, G. R. 1975, \apj, 196, 83 
\bibitem[Mitra \& Deshpande(1999)]{mit99} Mitra, D., \& Deshpande, A. A. 1999,
         \aap, preprint
\bibitem[Radhakrishnan \& Cooke (1969)]{rad69} Radhakrishnan, V., \& Cooke, D. J. 
        1969, \aplett, 3, 225
\bibitem[Rankin(1990)]{jra90} Rankin, J. M. 1990, \apj, 352, 247
\bibitem[Rankin(1993a)]{jr93a} Rankin, J. M. 1993a, \apj, 405, 285
\bibitem[Rankin(1993b)]{jr93b} Rankin, J. M. 1993b, \apjs, 85, 145
\bibitem[Ruderman \& Sutherland(1975)]{ras75} Ruderman, M. A., \& 
       Sutherland, P. G. 1975, \apj, 196, 51
\bibitem[Ruderman(1976)]{rud76} Ruderman, M. A. 1976, \apj, 203, 206
\bibitem[Sieber \& Oster(1975)]{sao75} Sieber, W., \& Oster, L. 1975, \aap, 38,
       325
\bibitem[Sieber \& Oster(1976)]{oas76} Sieber, W., \& Oster, L. 1976, \apj, 210,
       220
\bibitem[Suleymanova \& Izvekova(1984)]{sul84} Suleymanova, S. A., \& 
       Izvekova, V. A. 1984, Soviet Astronomy, 28, 32
\bibitem[Suleymanova et al.(1998)]{sul98} Suleymanova, S. A., Izvekova, V. A., 
        Rankin, J. M. \& Rathnasree, N. 1998, J. Astrophys. Astron., 19, 1  
\bibitem[Taylor \& Cordes(1993)]{tac93} Taylor, J. H. \& 
         Cordes, J. M. 1993, \apj, 401, 674
\bibitem[Taylor \& Huguenin(1971)]{tah71} Taylor, J. H., \& Huguenin, G. R. 
        1971, \apj, 167, 273
\bibitem[Taylor et al.(1971)]{tay71} Taylor, J. H., Huguenin, G. R., 
        Hirsch, R. M., \& Manchester, R. N. 1971, \aplett, 9, 205
\bibitem[Taylor et al.(1993)]{tay93} Taylor, J. H., Manchester, R. N., \& 
        Lyne, A. G.  1993, \apjs, 88, 529 
\bibitem[Usov \& Melrose(1995)]{uam95} Usov, V. V., \& Melrose, D. B. 1995,
        Australian J Phys., 48, 571
\bibitem[Vitkevich,  et al.(1969)] {vit69} Vitkevich, V. V., 
        Alexseev, Yu. I., \& Zhuravlev, Yu. P. 1969, Nature, 224, 49
\bibitem[Zhang \& Qiao(1996)]{zaq96} Zhang, B., \& Qiao, G. J. 1996, \aap, 310, 135
\bibitem[Zhang(1997)]{zha97} Zhang, B. 1997, \apj, 478, 313
\end{thebibliography}
\end{document}